\documentclass[aps,pra,twocolumn]{revtex4}
\usepackage{graphics}
\usepackage{amsmath}
\usepackage{graphicx}
\usepackage{amsfonts}
\usepackage{amssymb}
\usepackage{float}
\usepackage{longtable}
\usepackage{epsfig}
\usepackage{latexsym}
\usepackage{theorem}
\usepackage{bbm}
\usepackage{bm}
\usepackage{verbatim}
\usepackage{psfrag}
\newtheorem{theorem}{Theorem}

\newtheorem{remark}[theorem]{Remark}
\begin{document}
\title{General bounds for sender-receiver capacities in multipoint quantum communications}
\author{Riccardo Laurenza}
\affiliation{Computer Science and York Centre for Quantum Technologies, University of York,
York YO10 5GH, UK}
\author{Stefano Pirandola}
\affiliation{Computer Science and York Centre for Quantum Technologies, University of York,
York YO10 5GH, UK}

\begin{abstract}
We investigate the maximum rates for transmitting quantum information,
distilling entanglement and distributing secret keys between a sender and a
receiver in a multipoint communication scenario, with the assistance of
unlimited two-way classical communication involving all parties. First we
consider the case where a sender communicates with an arbitrary number of
receivers, so called quantum broadcast channel. Here we also provide a simple
analysis in the bosonic setting where we consider quantum broadcasting through
a sequence of beamsplitters. Then, we consider the opposite case where an
arbitrary number of senders communicate with a single receiver, so called
quantum multiple-access channel. Finally, we study the general case of
all-in-all quantum communication where an arbitrary number of senders
communicate with an arbitrary number of receivers. Since our bounds are
formulated for quantum systems of arbitrary dimension, they can be applied to
many different physical scenarios involving multipoint quantum communication.

\end{abstract}
\maketitle

\section{Introduction}

Today a huge effort is devoted to the development of robust quantum
technologies, directly inspired by the field of quantum
information~\cite{NielsenChuang,SamRMPm,RMP,hybrid,hybrid2,HolevoBOOK}. The
most typical communication tasks are quantum key distribution
(QKD)~\cite{BB84,Ekert,Gisin,Scarani,LM05,Gross,Chris,Twoway,Madsen,MDI1,CVMDIQKD}%
, reliable transmission of quantum information~\cite{QC1,QC2} and distribution
of entanglement~\cite{distillDV,B2main,distillCV}. The latter allows two
remote parties to implement powerful protocols such as quantum
teleportation~\cite{tele,teleCV,telereview}, which is a crucial tool for the
contruction of a future quantum Internet~\cite{Kimble2008,HybridINTERNET}.
Unfortunately, practical implementations are affected by
decoherence~\cite{Zurek}. This is the very reason why the performance of any
point-to-point protocol of quantum and private communication suffers from
fundamental limitations, which become more severe when the distance is
increased. This is the reason why we need quantum repeaters~\cite{repeaters}.

In this context, an open problem was to find the optimal rates for quantum and
private communication that are achievable by two remote parties, say Alice and
Bob, assuming the most general strategies allowed by quantum mechanics, i.e.,
assuming arbitrary local operations (LOs) assisted by unlimited two-way
classical communication (CCs), briefly called adaptive LOCCs. These optimal
rates are known as two-way (assisted) capacities and their determination has
been notoriously difficult. Only recently, after about $20$
years~\cite{ErasureChannelm}, Ref.~\cite{PLOB} finally addressed this problem
and established the two-way capacities at which two remote parties can
distribute entanglement ($D_{2}$), transmit quantum information ($Q_{2}$), and
generate secret keys ($K$) over a number of fundamental quantum channels at
both finite and infinite dimension, including erasure channels, dephasing
channels, bosonic lossy channels and quantum-limited amplifiers. For a review
of these results, see also Ref.~\cite{TQCreview}.

For the specific case of a bosonic lossy channel with transmissivity $\eta$,
Ref.~\cite{PLOB} proved that $D_{2}=Q_{2}=K=-\log_{2}(1-\eta)$ corresponding
to $\simeq1.44\eta$ bits per channel use at high loss. The latter result
completely characterizes the fundamental rate-loss scaling that affects any
point-to-point protocol of QKD through a lossy communication line, such as an
optical fiber or free-space link. The novel and general methodology that led
to these results is based on a suitable combination of quantum
teleportation~\cite{tele,teleCV,telereview} with a LOCC-monotonic functional,
such as the relative entropy of entanglement
(REE)~\cite{VedFORMm,Pleniom,RMPrelent}. Thanks to this combination,
Ref.~\cite{PLOB} was able to upper-bound the generic two-way capacity
$\mathcal{C}=D_{2}$, $Q_{2}$, $K$ of an arbitrary quantum channel
$\mathcal{E}$ with a computable single-letter quantity: This is the REE
$E_{R}(\sigma)$ of a suitable resource state $\sigma$ that is able to simulate
the quantum channel by means of a generalized teleportation protocol. In
particular, Ref.~\cite{PLOB} showed that $\sigma$ corresponds to the Choi
matrix of the channel when the channel is teleportation-covariant, i.e.,
suitably commutes with the random unitaries induced by the teleportation process.

The goal of the present paper is to extend such \textquotedblleft
REE+teleportation\textquotedblright\ methodology to a more complex
communication scenario, in particular that of a single-hop quantum network,
where multiple senders and/or receivers are involved. The basic configurations
are represented by the quantum broadcast channel~\cite{brd1,brd2,brd3} where
information is broadcast from a single sender to multiple receivers, and the
quantum multiple-access channel~\cite{mac2}, where multiple senders
communicate with a single receiver. More generally, we also consider the
combination of these two cases, where many senders communicate with many
receivers in a sort of all-in-all quantum communication or quantum
interference channel. In practical implementations, this may represent a
quantum bus~\cite{MaBus,BreBus} where quantum information is transmitted among
an arbitrary number of qubit registers.

In all these multipoint scenarios, we characterize the most general protocols
for entanglement distillation, quantum communication and key generation,
assisted by adaptive LOCCs. This leads to the definition of the two-way
capacities $\mathcal{C}=D_{2}$, $Q_{2}$, $K$ between any pair of sender and
receiver. We then consider those quantum channels (for broadcasting,
multiple-accessing, and all-in-all communication) which \text{are
teleportation-covariant}. For these channels, we can completely reduce an
adaptive protocol into a block form involving a tensor product of Choi
matrices. Combining this reduction with the REE, we then bound their two-way
capacities by means of the REE of their Choi matrix, therefore extending the
methods of Ref.~\cite{PLOB} to multipoint communication.

Our upper bounds applies to both discrete-variable (DV) and
continuous-variable (CV) channels. As an example, we consider the specific
case of a $1$-to-$M$ thermal-loss broadcast channel through a sequence of
beamsplitters subject to thermal noise. In particular, we discuss how that the
two-way capacities $Q_{2}$, $D_{2}$ and $K$ between the sender and each
receiver are all bounded by the first point-to-point channel in the
\textquotedblleft multisplitter\textquotedblright. This bottleneck result can
be extended to other Gaussian broadcast channels. In the specific case of a
lossy broadcast channel (without thermal noise), we find a straighforward
extension of the fundamental rate-loss scaling, so that any sender-receiver
capacity is bounded by $-\log_{2}(1-\eta)$ with $\eta$\ being the
transmissivity of the first beamsplitter.

The paper is organized as follows. In Sec.~\ref{review}, we review the basic
ideas, methods, and results of Ref.~\cite{PLOB} in relation to point-to-point
quantum and private communication. This serves as a background to the reader,
in order to better understand the novel developments which are presented in
the following sections about the multi-point communication. Specifically, we
consider the quantum broadcast channel in Sec.~\ref{SECbroadcasting}, the
quantum multiple-access channel in Sec.~\ref{SECmulti} and, finally, the
quantum interference channel in Sec.~\ref{SECall}. Sec.~\ref{SECconclu} is for conclusions.

\section{Theory of point-to-point quantum and private
communication\label{review}}

\begin{figure}[pth]
\vspace{-2.2cm}
\par
\begin{center}
\includegraphics[width=0.48\textwidth]{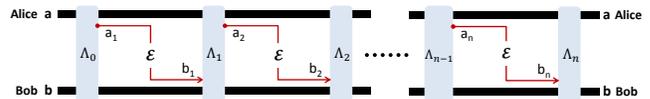} \vspace{-3.1cm}
\end{center}
\caption{Point-to-point adaptive protocol. Each transmission $a_{i}\rightarrow
b_{i}$ through the quantum channel $\mathcal{E}$ is interleaved by two
adaptive LOCCs, $\Lambda_{i-1}$ and $\Lambda_{i}$, applied to Alice's and
Bob's local registers $\mathbf{a}$ and $\mathbf{b}$. After $n$ transmissons,
Alice and Bob share an output state $\rho_{\mathbf{ab}}^{n}$.}%
\label{longPIC}%
\end{figure}

Let us briefly review the general methodology and results established in
Ref.~\cite{PLOB}. Let us define an adaptive point-to-point protocol through a
quantum channel $\mathcal{E}$. We assume that Alice has a local register
$\mathbf{a}$ (i.e., countable set of systems) and Bob has another local
register $\mathbf{b}$. These registers are prepared in some initial state
$\rho_{\mathbf{ab}}^{0}$ by means of an adaptive LOCC $\Lambda_{0}$. Then,
Alice picks a system $a_{1}\in\mathbf{a}$ and sends it through channel
$\mathcal{E}$; at the output, Bob gets a system $b_{1}$ which becomes part of
his register, i.e., $b_{1}\mathbf{b}\rightarrow\mathbf{b}$. Another adaptive
LOCC $\Lambda_{1}$ is applied to the registers. Then, there is the second
transmission $\mathbf{a}\ni a_{2}\rightarrow b_{2}$ through $\mathcal{E}$,
followed by another LOCC $\Lambda_{2}$ and so on (see Fig.~\ref{longPIC}).
After $n$ uses, Alice and Bob share an output state $\rho_{\mathbf{ab}}^{n}$
which is epsilon-close to a target state with $nR^{n}$ bits~\cite{Eps}. The
generic two-way capacity is defined by maximizing the asymptotic rate over all
the adaptive LOCCs $\mathcal{L}=\{\Lambda_{0},\ldots,\Lambda_{n}\}$, i.e.,%
\begin{equation}
\mathcal{C}(\mathcal{E}):=\sup_{\mathcal{L}}\lim_{n}R^{n}.
\end{equation}

In particular, by specifying the target state we identify a corresponding
two-way capacity. If the target state is a maximally-entangled state, then
$\mathcal{C}$ is equal to the two-way entanglement-distribution capacity
($D_{2}$). This is in turn equal to the two-way quantum capacity $Q_{2}$,
because the reliable transmission of a qubit implies the distribution of an
entanglement bit (ebit), and the distribution of an ebit implies the reliable
teleportation of a qubit. These operations are completely equivalent under
two-way CCs. If the target state is a private state~\cite{KD}, then
$\mathcal{C}$ is equal to the (two-way) secret key capacity ($K$). Since a
maximally-entangled state is a type of private state, we have $D_{2}\leq K$.
Then also note that $K=P_{2}$, where the latter is the two-way private
capacity of the channel, i.e., the maximum rate at which Alice may
deterministically transmit secret bits over the channel by means of adaptive
protocols~\cite{Devetak}. In summary, one has the hierarchy
\begin{equation}
D_{2}=Q_{2}\leq K=P_{2}. \label{hierarchy}%
\end{equation}

In order to bound these capacities by means of a computable single-letter
quantity, Ref.~\cite{PLOB} introduced a reduction\ method whose application
can be adapted to many other scenarios. The first ingredient is the extension
of the relative entropy of entanglement
(REE)~\cite{RMPrelent,VedFORMm,Pleniom} from quantum states to quantum
channels. Ref.~\cite[Theorem 1]{PLOB} showed that, for any quantum channel
$\mathcal{E}$ (at any dimension, finite or infinite), the generic two-way
capacity $\mathcal{C}(\mathcal{E})$ [i.e., any of the quantities in
Eq.~(\ref{hierarchy})] satisfies the weak converse bound%
\begin{equation}
\mathcal{C}(\mathcal{E})\leq E_{R}^{\bigstar}(\mathcal{E}):=\sup_{\mathcal{L}%
}\underset{n}{\lim}\frac{E_{R}(\rho_{\mathbf{ab}}^{n})}{n}~. \label{mainweak}%
\end{equation}

Here the adaptive channel's REE\ $E_{R}^{\bigstar}(\mathcal{E})$\ is defined
by computing the REE of the output state $\rho_{\mathbf{ab}}^{n}$, taking the
asymptotic limit in the number of channels uses $n$, and optimizing over all
the adaptive protocols $\mathcal{L}$. Recall the REE of a quantum state $\rho$
is defined as
\begin{equation}
E_{R}(\rho)=\inf_{\sigma_{s}}S(\rho||\sigma_{s}),
\end{equation}
where $\sigma_{s}$ is an arbitrary separable state and $S(\rho||\sigma
_{s}):=\mathrm{Tr}\left[  \rho(\log_{2}\rho-\log_{2}\sigma_{s})\right]  $ is
the relative entropy~\cite{RMPrelent}. For an asymptotic state $\sigma
:=\lim_{\mu}\sigma^{\mu}$ defined by a sequence of states $\sigma^{\mu}$, the
REE can be defined as~\cite{PLOB}
\begin{equation}
E_{R}(\sigma):=\inf_{\sigma_{s}^{\mu}}\underset{\mu\rightarrow+\infty}%
{\lim\inf}S(\sigma^{\mu}||\sigma_{s}^{\mu}), \label{REE_weaker}%
\end{equation}
where $\sigma_{s}^{\mu}$ is an arbitrary sequence of separable states which is
convergent in trace-norm, i.e., such that $||\sigma_{s}^{\mu}-\sigma
_{s}||\rightarrow0$ for some separable $\sigma_{s}$. This mathematical form is
directly inherited from the lower semi-continuity of the relative entropy for
CV\ systems~\cite{HolevoBOOK}.

\begin{remark}
The demonstration of Eq.~(\ref{mainweak}) exploits several tools from
Refs.~\cite{VedFORMm,Pleniom, KD,Donaldmain}. In particular, Ref.~\cite{PLOB}
provided three equivalent proofs, based on alternative treatments of the
private state involved in the definition of the secret key capacity. The first
proof exploits the fact that the dimension of the shield system~\cite{KD} of
the private state has an effective exponential scaling in the number of
channel uses; this scaling is an immediate application of well-known results
in the literature~\cite{Matthias1a,Matthias2a}, whose adaptation to CVs is
trivial as discussed in~\cite[Supplementary Note~3]{PLOB} and also in the
recent review~\cite{TQCreview}. The second proof assumes an exponential energy
growth in the channel uses, while the third proof does not depend on the
shield size. For a full discussion of these details see Supplementary Note~3
of Ref.~\cite{PLOB}.
\end{remark}

To simplify the upper bound of Eq.~(\ref{mainweak}) into a single-letter
quantity, Ref.~\cite{PLOB} devised a second ingredient. This consists of a
technique, dubbed \textquotedblleft teleportation stretching\textquotedblright%
, which reduces an adaptive protocol (with any communication task) into a much
simpler block-type protocol. More recently, this technique has been extended
to simplify adaptive protocols of quantum metrology and quantum channel
discrimination~\cite{Metro}. The first step of the technique is the LOCC
simulation of a quantum channel. This leads to a stretching of the channel
into a quantum state. Then, the second step is the exploitation of this
simulation argument in the adaptive protocol, so that all the transmissions
through the channel are replaced by a tensor product of quantum states subject
to a trace-preserving LOCC.

For any quantum channel $\mathcal{E}$, we may consider an LOCC-simulation.
This consists of an LOCC $\mathcal{T}$ and a resource state $\sigma$ such
that, for any input state $\rho$, the output of the channel can be expressed
as~\cite{PLOB}
\begin{equation}
\mathcal{E}(\rho)=\mathcal{T}(\rho\otimes\sigma). \label{sigma}%
\end{equation}
A channel $\mathcal{E}$ which is LOCC-simulable with a resource state $\sigma$
as in Eq.~(\ref{sigma}) is called \textquotedblleft$\sigma$%
-stretchable\textquotedblright~\cite{PLOB}. For the same channel $\mathcal{E}$
there may be different choices for $\mathcal{T}$ and $\sigma$. Furthermore,
the simulation can also be asymptotic. This means that we may consider
sequences of LOCCs $\mathcal{T}^{\mu}$ and resource states $\sigma^{\mu}$ such
that~\cite{PLOB}%
\begin{equation}
\mathcal{E}(\rho)=\lim_{\mu}\mathcal{T}^{\mu}(\rho\otimes\sigma^{\mu}),
\label{asymptotic}%
\end{equation}
for any input state $\rho$. In other words, a quantum channel may be defined
as a point-wise limit as in Eq.~(\ref{asymptotic}). This generalization is
relevant for the simulation of CV\ channels, such as bosonic channels, or
certain DV channels, such as the amplitude damping channel, whose\ simulation
is based on mappings between CVs and DVs.

\begin{remark}
The LOCC-simulation of an arbitrary quantum channel at any dimension (finite
or infinite) was introduced in Ref.~\cite{PLOB}. The first relevant idea was
the teleportation simulation of Ref.~\cite[Section~V]{B2main} whose
application was however limited to Pauli channels (as shown in
Ref.~\cite{SougatoBowen}). Other approaches known in the
literature~\cite{Niset,MHthesis,Wolfnotes,Leung} did not consider arbitrary
LOCC\ simulations but teleportation protocols, therefore restricting the
classes of simulable channels. In particular, the amplitude damping channel is
a simple example of a quantum channel that could not be deterministically
simulated by any approach prior to Ref.~\cite{PLOB}. Finally note that the
simulations in Refs.~\cite{Gatearray,Qsim0,Qsim1,Qsim2} are not suitable for
quantum communications because they generally imply non-local operations
between the remote parties. See Ref.~\cite[Supplementary Note~8]{PLOB} for
detailed discussions on the literature and advances in channel simulation. See
also the recent review~\cite{TQCreview} and Table~I therein.
\end{remark}

At any dimension (finite or infinite), an important class of quantum channels
are those \textquotedblleft Choi-stretchable\textquotedblright. These are
channels $\mathcal{E}$\ for which we may write the LOCC simulation of
Eqs.~(\ref{sigma}) with $\sigma$ being the Choi matrix of the channel, which
is defined as
\begin{equation}
\rho_{\mathcal{E}}:=\mathcal{I}\otimes\mathcal{E}(\Phi),
\end{equation}
where $\Phi$ is a maximally-entangled state. If the simulation is asymptotic,
then the Choi-stretchable channel satisfies Eq.~(\ref{asymptotic}) with
$\sigma^{\mu}$ being a sequence of Choi-approximating states $\rho
_{\mathcal{E}}^{\mu}$, i.e., such that their limit defines the asymptotic Choi
matrix of the channel as $\rho_{\mathcal{E}}:=\lim_{\mu}\rho_{\mathcal{E}%
}^{\mu}$. In particular, for a bosonic channel, we may set
\begin{equation}
\rho_{\mathcal{E}}^{\mu}:=\mathcal{I}\otimes\mathcal{E}(\Phi^{\mu}),
\end{equation}
where $\Phi^{\mu}$ is a two-mode squeezed vacuum (TMSV) state~\cite{RMP},
whose asymptotic limit defines the ideal CV\ Einstein-Podolsky-Rosen (EPR) state.

A simple criterion to identify Choi-stretchable channels is that of
teleportation-covariance. By definition, we say that a quantum channel
$\mathcal{E}$ is teleportation-covariant if, for any teleportation unitary $U$
(Pauli operators in DVs, phase-space displacements in CVs~\cite{telereview}%
),\ we may write%
\begin{equation}
\mathcal{E}(U\rho U^{\dagger})=V\mathcal{E}(\rho)V^{\dagger}~,
\label{stretchability}%
\end{equation}
for another (generally-different) unitary $V$~\cite{PLOB}. This is a large
family which includes Pauli, erasure and bosonic Gaussian channels. With this
definition in hand, Ref.~\cite[Proposition 2]{PLOB} showed that a
teleportation-covariant channel is certainly a Choi-stretchable channel, where
the LOCC\ simulation is simply given by quantum teleportation. In other words,
we may write $\mathcal{E}(\rho)=\mathcal{T}(\rho\otimes\rho_{\mathcal{E}})$
where $\mathcal{T}$ is a teleportation-LOCC. For asymptotic simulations, we
have
\begin{equation}
\mathcal{E}(\rho)=\lim_{\mu}\mathcal{T}^{\mu}(\rho\otimes\rho_{\mathcal{E}%
}^{\mu}), \label{asymptotics}%
\end{equation}
where $\mathcal{T}^{\mu}$\ is a sequence of teleportation-LOCCs. These are
built on finite-energy Gaussian measurements whose asymptotic limit defines
the ideal CV Bell detection~\cite{PLOB}.

\begin{remark}
The fact that teleportation-covariance implies the simulation of the channel
by means of teleportion over the channel's Choi matrix has been first
discussed in Ref.~\cite{Leung}\ in the context of DV channels. It has been
later generalized in Ref.~\cite{PLOB} to include CV channels and asymptotic
simulations. See also Ref.~\cite{TQCreview}.
\end{remark}

\begin{figure}[pth]
\vspace{-0.5cm}
\par
\begin{center}
\includegraphics[width=0.48\textwidth]{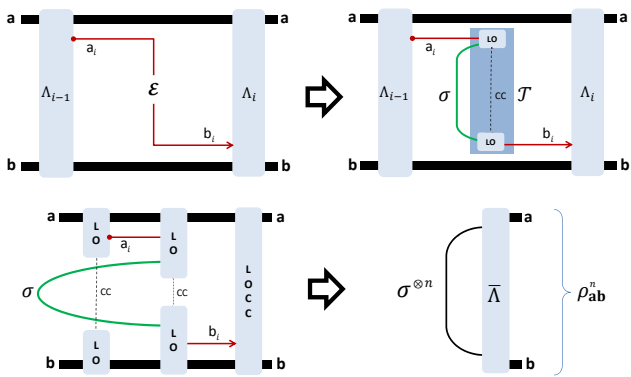} \vspace{-1.2cm}
\end{center}
\caption{Teleportation stretching of an adaptive point-to-point protocol.
\textbf{(Top panels)~}The generic $i$th transmission through the channel
$\mathcal{E}$ is simulated by an LOCC $\mathcal{T}$\ and a resource state
$\sigma$. (\textbf{Bottom panels}) The resource state $\sigma$ is stretched
back in time out of the adaptive LOCCs. This is repeated for all the $n$
transmissions, so that we accumulate the tensor-product state $\sigma^{\otimes
n}$. All the LOCCs (those of the original protocol and those introduced by the
simulation) are collapsed into a single trace-preserving LOCC $\bar{\Lambda}$
(which also includes the initial state of the registers $\rho_{\mathbf{ab}%
}^{0}$).}%
\label{pppPIC}%
\end{figure}

Thanks to the LOCC-simulation $(\mathcal{T},\sigma)$ of a quantum channel
$\mathcal{E}$ as in Eq.~(\ref{sigma}), one may completely simplify the
structure of an adaptive protocol. In fact, the output $\rho_{\mathbf{ab}}%
^{n}$ can be reduced to a tensor-product $\sigma^{\otimes n}$ up to a
trace-preserving LOCC $\bar{\Lambda}$~\cite{PLOB}. In other words, we may
write
\begin{equation}
\rho_{\mathbf{ab}}^{n}=\bar{\Lambda}\left(  \sigma^{\otimes n}\right)  ~.
\label{StretchingMAIN}%
\end{equation}
In fact, (i)~first we replace each transmission through the channel
$\mathcal{E}$ with an LOCC-simulation $(\mathcal{T},\sigma)$; (ii)~then we
stretch the resource state $\sigma$ \textquotedblleft back in
time\textquotedblright; (iii)~finally, we collapse all the LOCCs (and also the
initial separable state $\rho_{\mathbf{ab}}^{0}$) into a single
trace-preserving LOCC (which is suitably averaged over all the possible
measurements in the simulated protocol). These steps are depticted in
Fig.~\ref{pppPIC} and lead to the decomposition in Eq.~(\ref{StretchingMAIN}%
).

For a quantum channel with asymptotic simulation as in Eq.~(\ref{asymptotic}),
the procedure is more involved. One first considers an imperfect channel
simulation $\mathcal{E}^{\mu}(\rho):=\mathcal{T}^{\mu}(\rho\otimes\sigma^{\mu
})$ in each transmission. By adopting this simulation, we realize an imperfect
stretching of the protocol, with output state $\rho_{\mathbf{ab}}^{\mu
,n}:=\bar{\Lambda}_{\mu}\left(  \sigma^{\mu\otimes n}\right)  $\ for a
trace-preserving LOCC $\bar{\Lambda}_{\mu}$. This is done similarly to the
steps in Fig.~\ref{pppPIC}, but considering $\mathcal{E}^{\mu}$ in the place
of the original channel $\mathcal{E}$. A crucial point is now the estimation
of the error in the channel simulation, which must be suitably controlled and
propagated to the output state.

Assume that the registers have a total number $m$ of modes (the value of $m$
can be taken to be finite and then relaxed at the very end to include
countable registers). Then assume that, during the $n$ transmissions of the
protocol, the total mean number of photons in the registers is bounded by some
large but finite value $N$. We may therefore define the set of
energy-constrained states
\begin{equation}
\mathcal{D}_{N}:=\{\rho_{\mathbf{ab}}~|~\mathrm{Tr}(\hat{N}\rho_{\mathbf{ab}%
})\leq N\}\subset\mathcal{D}(\mathcal{H}^{\otimes m}),
\end{equation}
where $\hat{N}$ is the $m$-mode number operator. For the $i$th transmission
$a_{i}\mathbf{\rightarrow}b_{i}$, the simulation error\ may be quantified in
terms of the energy-bounded diamond norm~\cite{PLOB}
\begin{align}
\varepsilon_{N}  &  :=\left\Vert \mathcal{E}-\mathcal{E}^{\mu}\right\Vert
_{\diamond N}=\\
&  \sup_{\rho_{a_{i}\mathbf{ab}}\in\mathcal{D}_{N}}\left\Vert \mathcal{E}%
\otimes\mathcal{I}_{\mathbf{ab}}(\rho_{a_{i}\mathbf{ab}})-\mathcal{E}^{\mu
}\otimes\mathcal{I}_{\mathbf{ab}}(\rho_{a_{i}\mathbf{ab}})\right\Vert
.\nonumber
\end{align}

Because $\mathcal{D}_{N}$ is compact~\cite{HolevoCOMPACT} and channel
$\mathcal{E}$ is defined by the point-wise limit $\mathcal{E}(\rho)=\lim_{\mu
}\mathcal{E}^{\mu}(\rho)$, we may write the following uniform limit
\begin{equation}
\varepsilon_{N}\overset{\mu}{\rightarrow}0\mathrm{~~}\text{for~any }N.
\end{equation}
This error has to be propagated to the output state, so that we can suitably
bound the trace distance between the actual output $\rho_{\mathbf{ab}}^{n}$
and the simulated output $\rho_{\mathbf{ab}}^{n,\mu}$. By using basic
properties of the trace distance (triangle inequality and monotonicity under
maps), Ref.~\cite{PLOB} showed that the simulation error in the output state
satisfies%
\begin{equation}
\left\Vert \rho_{\mathbf{ab}}^{n}-\rho_{\mathbf{ab}}^{n,\mu}\right\Vert \leq
n\left\Vert \mathcal{E}-\mathcal{E}^{\mu}\right\Vert _{\diamond N}~.
\end{equation}
Therefore, for any $N$, we may write the trace-norm limit
\begin{equation}
\left\Vert \rho_{\mathbf{ab}}^{n}-\bar{\Lambda}_{\mu}\left(  \sigma
^{\mu\otimes n}\right)  \right\Vert \overset{\mu}{\rightarrow}0,
\label{stretch2}%
\end{equation}
i.e., the asymptotic stretching $\rho_{\mathbf{ab}}^{n}=\lim_{\mu}\bar
{\Lambda}_{\mu}(\sigma^{\mu\otimes n})$.

\begin{remark}
Teleportation stretching simplifies an arbitrary adaptive protocol implemented
over an arbitrary channel at any dimension, finite or infinite.\ In
particular, it works by maintaining the original communication task. This
means that\ an adaptive protocol of quantum communication (QC), entanglement
distribution (ED) or key generation (KG), is reduced to a corresponding block
protocol with exactly the same original task (QC, ED, or KG), but with the
output state being decomposed in the form of Eq.~(\ref{StretchingMAIN}) or
Eq.~(\ref{stretch2}). In the literature, there were some precursory but
restricted arguments, as those in Refs.~\cite{B2main,Niset}. These were
limited to the transformation of a protocol of QC into a protocol of ED, over
specific classes of channels (e.g., Pauli channels in Ref.~\cite{B2main}).
Furthermore, no control of the simulation error was considered in previous
literature~\cite{Niset}, while this is crucial for the rigorous simulation of
bosonic channels.
\end{remark}

The most crucial insight of Ref.~\cite{PLOB} has been the combination of the
previous two ingredients, i.e., channel's REE and teleportation stretching,
which is the key observation leading to a single-letter upper bound for all
the two-way capacities of a quantum channel. In fact, let us compute the
REE\ of the output state decomposed as in Eq.~(\ref{StretchingMAIN}). We
derive%
\begin{equation}
E_{R}(\rho_{\mathbf{ab}}^{n})\overset{(1)}{\leq}E_{R}(\sigma^{\otimes
n})\overset{(2)}{\leq}nE_{R}(\sigma)~, \label{toREP}%
\end{equation}
using (1) the monotonicity of the REE under trace-preserving LOCCs and (2) its
subadditive over tensor products. By replacing Eq.~(\ref{toREP}) in
Eq.~(\ref{mainweak}), we then find the single-letter upper
bound~\cite[Theorem~5]{PLOB}
\begin{equation}
\mathcal{C}(\mathcal{E})\leq E_{R}(\sigma)~. \label{UB1}%
\end{equation}
In particular, if the channel $\mathcal{E}$ is teleportation-covariant, it is
Choi-stretchable, and we may write~\cite[Theorem 5]{PLOB}
\begin{equation}
\mathcal{C}(\mathcal{E})\leq E_{R}(\rho_{\mathcal{E}}). \label{UB2}%
\end{equation}

These results are suitable extended to asymptotic simulations. By adopting the
extended definition of REE in Eq.~(\ref{REE_weaker}), Ref.~\cite{PLOB} showed
that Eqs.~(\ref{UB1}) and~(\ref{UB2}) are valid for channels with asymptotic
simulations, such as bosonic channels. In particular, the proof exploits the
fact that Eq.~(\ref{mainweak}) involves a supremum over all protocols
$\mathcal{L}$, so that we may extend the upper bound to the asymptotic limit
of energy-uncostrained protocols where the total mean photon number $N$ tends
to infinite (and the local registers become countable set).

The upper bound of Eq.~(\ref{UB2}) is valid for any teleportation-covariant
channel, in particular for Pauli channels (e.g., depolarizing or dephasing),
erasure channels and bosonic Gaussian channels. Then, by showing coincidence
of this upper bound with lower bounds based on the coherent~\cite{QC1,QC2} and
reverse coherent information~\cite{RevCohINFO,ReverseCAP}, Ref.~\cite{PLOB}
established strikingly simple formulas for the two-way capacities of the most
fundamental quantum channels. For a bosonic lossy channel $\mathcal{E}_{\eta}%
$\ with transmissivity $\eta$~\cite{RMP}, one has~\cite{PLOB}
\begin{equation}
D_{2}(\eta)=Q_{2}(\eta)=K(\eta)=P_{2}(\eta)=-\log_{2}(1-\eta)~.
\label{formCloss}%
\end{equation}
In particular, the secret-key capacity\ of the lossy channel determines the
maximum rate achievable by any QKD protocol. At high loss $\eta\simeq0$, one
has the optimal rate-loss scaling of $K\simeq1.44\eta$ secret bits per channel
use. Because it establishes the upper limit of any point-to-point quantum
optical communication, Eq.~(\ref{formCloss}) also establishes a
\textquotedblleft repetearless bound\textquotedblright, i.e., the benchmark
that quantum repeaters must surpass in order to be effective.

Then, for a quantum-limited amplifier $\mathcal{E}_{g}$ with gain
$g>1$~\cite{RMP}, one finds~\cite{PLOB}%
\begin{equation}
D_{2}(g)=Q_{2}(g)=K(g)=P_{2}(g)=-\log_{2}\left(  1-g^{-1}\right)  ~.
\label{Campli}%
\end{equation}
In particular, this proves that $Q_{2}(g)$ coincides with the unassisted
quantum capacity $Q(g)$~\cite{HolevoWerner,Wolf}. For a qubit dephasing
channel $\mathcal{E}_{p}^{\text{deph}}$ with dephasing probability $p$, one
has~\cite{PLOB}%
\begin{equation}
D_{2}(p)=Q_{2}(p)=K(p)=P_{2}(p)=1-H_{2}(p)~, \label{dep2}%
\end{equation}
where $H_{2}$ is the binary Shannon entropy. Note that this also proves
$Q_{2}(p)=Q(p)$ for a dephasing channel, where $Q(p)$ was found in
Ref.~\cite{degradable}. Eq.~(\ref{dep2}) can be extended to dephasing channels
$\mathcal{E}_{p,d}^{\text{deph}}$ in arbitrary dimension $d$, so that all the
two-way capacities are given by~\cite{PLOB}
\begin{equation}
\mathcal{C}(p,d)=\log_{2}d-H(\{P_{i}\})~,
\end{equation}
where $H$ is the Shannon entropy and $P_{i}$ is the probability of $i$ phase
flips. Finally, for the qudit erasure channel $\mathcal{E}_{p,d}%
^{\text{erase}}$ with erasure probability $p$, one finds~\cite{PLOB}
\begin{equation}
D_{2}(p)=Q_{2}(p)=K(p)=P_{2}(p)=(1-p)\log_{2}d~. \label{erase2}%
\end{equation}
As previously mentioned, only the $Q_{2}$ of the erasure channel was
previously known~\cite{ErasureChannelm}. Simultaneously with Ref.~\cite{PLOB},
an independent study of the erasure channel has been provided by
Ref.~\cite{GEWa} which showed how its $K$ can be computed from the squashed
entanglement (see also Ref.~\cite[Supplementary~Discussion (page 38)]{PLOB}).

\section{Quantum broadcast channel\label{SECbroadcasting}}

Here we consider quantum and private communication in a single-hop
point-to-multipoint network. We adapt the techniques of Ref.~\cite{PLOB} to
bound the optimal rates that are achievable in adaptive protocols involving
multiple receivers. For the sake of simplicity, we present the theory for
non-asymptotic simulations. The theoretical treatment of asymptotic
simulations goes along the lines described in previous Sec.~\ref{review} and
is discussed afterwards.\

Consider a quantum broadcast channel $\mathcal{E}$\ where Alice (local
register $\mathbf{a}$) transmits a system $a\in\mathbf{a}$ to $M$ different
Bobs; the generic $i$th Bob (with $i=1,\ldots,M$) receives an output system
$b^{i}$ which may be combined with a local register $\mathbf{b}^{i}$ for
further processing. Denote by $\mathcal{D}(\mathcal{H}_{s})$ the space of
density operators defined over the Hilbert space $\mathcal{H}_{s}$ of quantum
system $s$. Then, the quantum broadcast channel is a completely-positive trace
preserving (CPTP) map from Alice's input space $\mathcal{D}(\mathcal{H}_{a})$
to the Bobs' output space $\mathcal{D}(\otimes_{i}\mathcal{H}_{b^{i}})$. The
most general adaptive protocol over this channel goes as follows.

All the parties prepare their initial systems by means of a LOCC $\Lambda_{0}%
$. Then, Alice picks the first system $a_{1}\in\mathbf{a}$ which is broadcast
to all Bobs $a_{1}\rightarrow\{b_{1}^{i}\}$ through channel $\mathcal{E}$.
This is followed by another LOCC $\Lambda_{1}$ involving all parties. Bobs'
ensembles are updated as $b_{1}^{i}\mathbf{b}^{i}\rightarrow\mathbf{b}^{i}$.
Then, there is the second broadcast $\mathbf{a}\ni a_{2}\rightarrow\{b_{2}%
^{i}\}$ through $\mathcal{E}$, followed by another LOCC $\Lambda_{2}$ and so
on. After $n$ uses, Alice and the $i$th Bob share an output state
$\rho_{\mathbf{ab}^{i}}^{n}$ which is epsilon-close to a target state with
$nR_{i}^{n}$ bits. The generic broadcast capacity for the $i$th Bob is defined
by maximizing the asymptotic rate over all the adaptive LOCCs $\mathcal{L}%
=\{\Lambda_{0},\Lambda_{1},\ldots\}$, i.e., we have
\begin{equation}
\mathcal{C}^{i}:=\sup_{\mathcal{L}}\lim_{n}R_{i}^{n}.
\end{equation}

By specifying the adaptive protocol to a particular target state, i.e., to a
particular task (entanglement distribution, reliable transmission of quantum
information, key generation or deterministic transmission of secret bits), one
derives the entanglement-distribution broadcast capacity ($D_{2}^{i}$), the
quantum broadcast capacity ($Q_{2}^{i}$), the secret-key broadcast capacity
($K^{i}$), and the private broadcast capacity ($P_{2}^{i}$). These are all
assisted by unlimited two-way CCs between the parties and it is easy to check
that they must satisfy $D_{2}^{i}=Q_{2}^{i}\leq K^{i}=P_{2}^{i}$.

\begin{figure}[ptbh] \vspace{-2.2cm}
\par
\begin{center}
\includegraphics[width=0.60\textwidth]{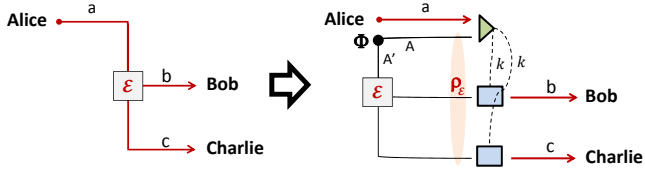} \vspace{-3.4cm}
\end{center}
\caption{Simulation of a teleportation-covariant quantum broadcast channel. We
may replace the broadcast channel $\mathcal{E}:a\rightarrow bc$\ by
teleportation over its Choi matrix $\rho_{\mathcal{E}}$, with CCs to Bob and
Charlie, who will implement correction unitaries. The broadcast channel is
therefore Choi-stretchable and its LOCC simulation is based on teleportation.}%
\label{broad}%
\end{figure}

In order to bound the previous capacities, let us introduce the notion of
teleportation-covariant broadcast channel. It is explained for the case of two
receivers, Bob and Charlie, with the extension to arbitrary $M$ receivers
being just a matter of technicalities. This is a broadcast channel which
suitably commutes with teleportation. Formally, this means that, for any
teleportation unitary $U_{k}$ at the channel input, we may write
\begin{equation}
\mathcal{E}(U_{k}\rho U_{k}^{\dagger})=(B_{k}\otimes C_{k})\mathcal{E}%
(\rho)(B_{k}\otimes C_{k})^{\dagger}~,
\end{equation}
for unitaries $B_{k}$ and $C_{k}$ at the two outputs. If this is the
case,\textbf{ }it is immediate to prove that $\mathcal{E}$ can be simulated by
a generalized teleportation protocol over its Choi matrix
\begin{equation}
\rho_{\mathcal{E}}=\mathcal{I}_{A}\otimes\mathcal{E}_{A^{\prime}}%
(\Phi_{AA^{\prime}}), \label{choi1}%
\end{equation}
where the latter is defined by sending half of an EPR
$\Phi_{AA^{\prime}}$ through the broadcast channel. In other
words, the broadcast channel is Choi-stretchable and its LOCC
simulation is based on teleportation. See
Fig.~\ref{broad}.

\begin{figure}[ptbh]
\vspace{-0.3cm}
\par
\begin{center}
\includegraphics[width=0.50\textwidth]{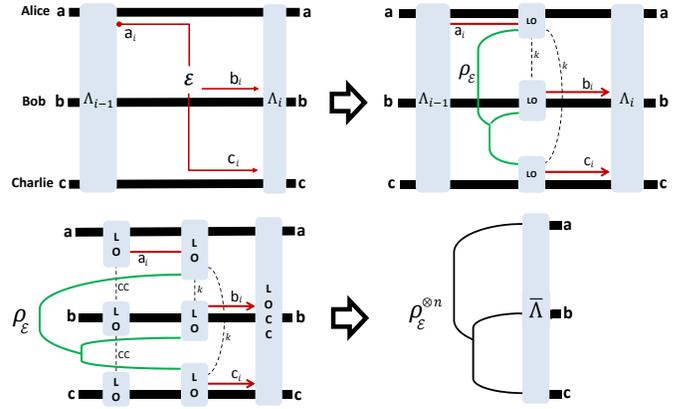} \vspace{-1.1cm}
\end{center}
\caption{Stretching an adaptive protocol over a teleportation-covariant
quantum broadcast channel. {\bfseries Top panels}. The generic $i$th
transmission $a_{i}\rightarrow\{b_{i},c_{i}\}$ over the broadcast channel
$\mathcal{E}$ (red line) is replaced by a teleportation over its Choi matrix
$\rho_{\mathcal{E}}$ (following the procedure shown in Fig.~\ref{broad}). The
input system and the upper half of the Choi matrix are subject to a Bell
detection which becomes part of Alice's LO (upper LO). The result of the Bell
detection $k$ is classically communicated to Bob and Charlie so that they can
apply two correction unitaries which are then included into their respective
LOs (middle and lower LOs). {\bfseries Bottom panels}. The Choi matrix is
stertched in time out of the adaptive LOCCs which are then collapsed into a
single trace-preserving LOCC. After $n$ uses, we can express the output in
terms of $n$ copies of the Choi matrix $\rho_{\mathcal{E}}$ of the broadcast
channel, plus a trace-preserving single final LOCC $\bar{\Lambda}$ as in
Eq.~(\ref{tensorOUT}).}%
\label{broadST}%
\end{figure}

Following and extending the ideas of Ref.~\cite{PLOB}, we may simplify any
adaptive protocol performed over a teleportation-covariant broadcast channel.
The steps of the procedure are shown in Fig.~\ref{broadST}. As a result, the
total output state of Alice, Bob and Charlie can be decomposed in the form%
\begin{equation}
\rho_{\mathbf{abc}}^{n}:=\rho_{\mathbf{abc}}(\mathcal{E}^{\otimes n}%
)=\bar{\Lambda}\left(  \rho_{\mathcal{E}}^{\otimes n}\right)  ~,
\label{tensorOUT}%
\end{equation}
where $\bar{\Lambda}$\ is a trace-preserving LOCC. If we now trace one of the
two receivers, e.g., Charlie, we still have a trace-preserving LOCC between
Alice and Bob, and we may write the following
\begin{equation}
\rho_{\mathbf{ab}}^{n}=\mathrm{Tr}_{\mathbf{c}}\bar{\Lambda}\left(
\rho_{\mathcal{E}}^{\otimes n}\right)  =\bar{\Lambda}_{\mathbf{a|bc}}\left(
\rho_{\mathcal{E}}^{\otimes n}\right)  ~, \label{broadppp}%
\end{equation}
where $\bar{\Lambda}_{\mathbf{a|bc}}$ is local with respect to the cut
$\mathbf{a|bc}$.

Let us now compute the REE\ of Alice and Bob's output state $\rho
_{\mathbf{ab}}^{n}$. Using Eq.~(\ref{broadppp}) and the monotonicity of the
REE under $\bar{\Lambda}_{\mathbf{a|bc}}$, we derive%
\begin{align}
E_{R}(\rho_{\mathbf{ab}}^{n})  &  :=\inf_{\sigma_{s}(\mathbf{a}|\mathbf{b}%
)}S\left(  \rho_{\mathbf{ab}}^{n}||\sigma_{s}\right) \nonumber\\
&  \leq\inf_{\sigma_{s}(\mathbf{a}|\mathbf{bc})}S\left(  \rho_{\mathcal{E}%
}^{\otimes n}||\sigma_{s}\right)  :=E_{R(\mathbf{a}|\mathbf{bc})}%
(\rho_{\mathcal{E}}^{\otimes n}), \label{ooo1}%
\end{align}
where we call $E_{R(\mathbf{a}|\mathbf{bc})}$ the REE\ with respect to the
bipartite cut $\mathbf{a}|\mathbf{bc}$~\cite{monotonocity}. Note that the set
of states $\{\sigma_{s}(\mathbf{a}|\mathbf{bc})\}$, separable between
$\mathbf{a}$ and $\mathbf{bc}$, includes the set of states $\{\sigma
_{s}(\mathbf{a}|\mathbf{b}|\mathbf{c})\}$ which are separable with respect to
$\mathbf{a}$, $\mathbf{b}$ and $\mathbf{c}$. Therefore, we may write the
further upper-bound
\begin{equation}
E_{R(\mathbf{a}|\mathbf{bc})}(\rho_{\mathcal{E}}^{\otimes n})\leq\inf
_{\sigma_{s}(\mathbf{a}|\mathbf{b}|\mathbf{c})}S\left(  \rho_{\mathcal{E}%
}^{\otimes n}||\sigma_{s}\right)  :=E_{R}(\rho_{\mathcal{E}}^{\otimes n}).
\label{ooo2}%
\end{equation}

For Alice and Bob ($i=B$), we can then exploit the weak converse bound in
Eq.~(\ref{mainweak}) where the optimization must be done over all the adaptive
broadcast protocols. Combining this bound with Eqs.~(\ref{ooo1})
and~(\ref{ooo2}), we get%
\begin{equation}
\mathcal{C}^{B}\leq\sup_{\mathcal{L}}\underset{n}{\lim}\frac{E_{R}%
(\rho_{\mathbf{ab}}^{n})}{n}\leq E_{R(\mathbf{a}|\mathbf{bc})}^{\infty}%
(\rho_{\mathcal{E}})\leq E_{R}^{\infty}(\rho_{\mathcal{E}}),
\end{equation}
where $E_{R}^{\infty}(\rho):=\lim_{n}n^{-1}E_{R}(\rho^{\otimes n})$ is the
regularized version of the REE. Then, using the subadditive over tensor
products, we may also write
\begin{equation}
E_{R}(\rho_{\mathbf{ab}}^{n})\leq nE_{R(\mathbf{a}|\mathbf{bc})}%
(\rho_{\mathcal{E}})\leq nE_{R}(\rho_{\mathcal{E}}),
\end{equation}
which clearly leads to the single-letter upper bounds
\begin{equation}
\mathcal{C}^{B}\leq E_{R(\mathbf{a}|\mathbf{bc})}(\rho_{\mathcal{E}})\leq
E_{R}(\rho_{\mathcal{E}}). \label{ERRdv}%
\end{equation}

We find the same bounds for the capacity of Alice and Charlie ($i=C$). In
general, for arbitrary $M$ receivers, we may extend the reasoning and write
the following upper bounds for the capacity between Alice and the $i$th Bob%
\begin{equation}
\mathcal{C}^{i}\leq E_{R(\mathbf{a}|\mathbf{b}^{1}\cdots\mathbf{b}^{M})}%
(\rho_{\mathcal{E}})\leq E_{R}(\rho_{\mathcal{E}}):=\Phi(\mathcal{E})~,
\label{upper}%
\end{equation}
where $\Phi(\mathcal{E})$\ is the entanglement flux of the broadcast channel
$\mathcal{E}$, defined as the REE of its Choi matrix $\rho_{\mathcal{E}}$.

\subsection{Extension to continuous variables\label{SecCV}}

As explained in Sec.~\ref{review}, one cannot directly apply the DV
formulation of channel simulation and teleportation stretching to CV systems.
There are non-trivial issues to be taken into account, related with the
infinite energy of the asymptotic Choi matrices of the bosonic channels. These
issues require a suitable treatment~\cite{PLOB,TQCreview}.

The Choi matrix of a bosonic broadcast channel can be defined as the following
asymptotic state%
\begin{equation}
\rho_{\mathcal{E}}:=\lim_{\mu}\rho_{\mathcal{E}}^{\mu},~~~\rho_{\mathcal{E}%
}^{\mu}=\mathcal{I}_{A}\otimes\mathcal{E}_{A^{\prime}}(\Phi_{AA^{\prime}}%
^{\mu}),
\end{equation}
with $\Phi_{AA^{\prime}}^{\mu}$ being a TMSV state. The simulation of a
teleportation-covariant bosonic broadcast channel is based on the sequence of
Choi-approximating states $\rho_{\mathcal{E}}^{\mu}$, so that we may write the
generalization of Eq.~(\ref{asymptotics}), i.e.,
\begin{equation}
\mathcal{E}(\rho)=\lim_{\mu}\mathcal{T}^{\mu}(\rho\otimes\rho_{\mathcal{E}%
}^{\mu}),
\end{equation}
where $\mathcal{T}^{\mu}$\ is a sequence of teleportation-LOCCs. By repeating
the reasoning in Sec.~\ref{review}, the error in the channel simulation can be
propagated to the output state of the adaptive protocol, so that, for any
energy constraint on the local registers, we may write the trace-norm limit
\begin{equation}
\left\Vert \rho_{\mathbf{abc}}^{n}-\bar{\Lambda}_{\mu}\left(  \rho
_{\mathcal{E}}^{\mu\otimes n}\right)  \right\Vert \overset{\mu}{\rightarrow}0,
\end{equation}
where $\bar{\Lambda}_{\mu}$\ is an imperfect stretching-LOCC associated with
the imperfect teleportation LOCC $\mathcal{T}^{\mu}$. By tracing one of the
outputs, e.g., Charlie, one gets
\begin{equation}
\left\Vert \rho_{\mathbf{ab}}^{n}-\bar{\Lambda}_{\mu}^{\mathbf{a|bc}}\left(
\rho_{\mathcal{E}}^{\mu\otimes n}\right)  \right\Vert \overset{\mu
}{\rightarrow}0, \label{stttt}%
\end{equation}
where $\bar{\Lambda}_{\mu}^{\mathbf{a|bc}}$ is an imperfect stretching-LOCC
associated with Alice and Bob, which is local with respect to the bipartite
cut $\mathbf{a|bc}$.

The next step is to extend the definition of REE to asymptotic states as in
Eq.~(\ref{REE_weaker}). In particular, we define%
\begin{equation}
E_{R(\mathbf{a|bc})}(\rho_{\mathcal{E}}):=\inf_{\sigma_{s}^{\mu}%
(\mathbf{a|bc})}\underset{\mu\rightarrow+\infty}{\lim\inf}S(\rho_{\mathcal{E}%
}^{\mu}||\sigma_{s}^{\mu}), \label{broadFFLL}%
\end{equation}
where $\sigma_{s}^{\mu}(\mathbf{a|bc})$ is an arbitrary converging sequence of
states that is separable with respect to the cut $\mathbf{a|bc}$. Then, we
also define the entanglement flux of the bosonic broadcast channel as%
\begin{equation}
\Phi(\mathcal{E})=E_{R}(\rho_{\mathcal{E}}):=\inf_{\sigma_{s}^{\mu}}%
\underset{\mu\rightarrow+\infty}{\lim\inf}S(\rho_{\mathcal{E}}^{\mu}%
||\sigma_{s}^{\mu}), \label{broadFFFLLL}%
\end{equation}
where $\sigma_{s}^{\mu}$ is an arbitrary converging sequence of separable
states (with respect to all the local systems $\mathbf{a|b|c}$). By applying a
direct extension of the weak converse bound in Eq.~(\ref{mainweak}), we then
derive the same result as in Eq.~(\ref{ERRdv}) for the capacity $\mathcal{C}%
^{B}$ between Alice and Bob, proviso that the REE\ quantities are suitably
extended as in Eqs.~(\ref{broadFFLL}) and~(\ref{broadFFFLLL}). In general, for
arbitrary $M$ receivers, we have the corresponding extension of
Eq.~(\ref{upper}).

\subsection{Thermal-loss quantum broadcast channel}

Now that we have rigorously extended the treatment to CV systems, we study the
example of a bosonic broadcast channel from Alice to $M$ Bobs which introduces
both loss and thermal noise. This is an optical scenario that may easily occur
in practice. For instance, it may represent the practical implementation of a
single-hop QKD\ network, where a party wants to share keys with several other
parties for broadcasting private information. The latter may also be a common
key to enable a quantum-secure conferencing among all the trusted parties.

\begin{figure}[ptbh]
\vspace{-1.5cm}
\par
\begin{center}
\vspace{-0.5cm} \vspace{0.5cm} \vspace{-0.7cm}
\includegraphics[width=0.59\textwidth]{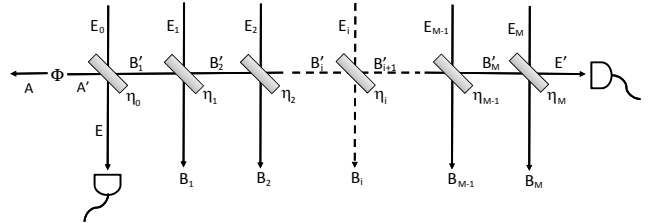} \vspace{-0.5cm}
\vspace{-2.3cm}
\end{center}
\caption{Thermal-loss quantum broadcast channel $\mathcal{E}_{A^{\prime
}\rightarrow B_{1}\ldots B_{M}}$\ from Alice (mode $A^{\prime}$) to $M$ Bobs
(modes $B_{1},\ldots,B_{M}$), realized by a multi-splitter, i.e., a sequence
of $M+1$ beamsplitters with transmissivities $(\eta_{0},\eta_{1},\ldots
\eta_{M})$. The environmental modes $E_{0},E_{1}\ldots,E_{M}$ are in thermal
states. Modes $E$ and $E^{\prime}$ describe leakage to the environment.}%
\label{broadBS}%
\end{figure}

One possible physical representation is a chain of $M+1$ beamsplitters with
transmissivities $(\eta_{0},\eta_{1},\ldots\eta_{M})$ in which Alice's input
mode $A^{\prime}$ subsequently interacts with $M+1$ modes $(E_{0},E_{1}%
,E_{2},\ldots,E_{M})$ described by thermal states $\rho_{E_{i}}(\bar{n}_{i})$
with $\bar{n}_{i}$ mean number of photons. The $M$ output modes $(B_{1}%
,B_{2},\ldots,B_{M})$ are then given to the different Bobs, with the extra
modes $E$ and $E^{\prime}$\ being the leakage to the environment (or an
eavesdropper). See Fig.~\ref{broadBS} for a schematic representation of this
thermal-loss broadcast channel $\mathcal{E}=\mathcal{E}_{A^{\prime}\rightarrow
B_{1}\ldots B_{M}}$.

The generic capacity $\mathcal{C}^{i}$ between Alice and the $i$th Bob is
upper bounded by%
\begin{equation}
\mathcal{C}^{i}\leq E_{R(A|B_{1}\cdots B_{M})}(\rho_{\mathcal{E}}%
):=\inf_{\sigma_{s}^{\mu}(A|B_{1}\cdots B_{M})}\underset{\mu\rightarrow
+\infty}{\lim\inf}S(\rho_{\mathcal{E}}^{\mu}||\sigma_{s}^{\mu}),
\end{equation}
where the state $\rho_{\mathcal{E}}^{\mu}:=\mathcal{I}_{A}\otimes
\mathcal{E}_{A^{\prime}\rightarrow B_{1}\ldots B_{M}}(\Phi_{AA^{\prime}}^{\mu
})$ is the Choi-approximating state obtained by sending one half of a TMSV
state $\Phi_{AA^{\prime}}^{\mu}$, and $\sigma_{s}^{\mu}(A|B_{1}\cdots B_{M})$
is a converging sequence of states that are separable with respect to the cut
$A|B_{1}\cdots B_{M}$. Now notice that we may write
\begin{equation}
\rho_{\mathcal{E}}^{\mu}=\mathbf{L}_{A|B_{1}^{\prime}E_{1}\cdots E_{M}}\left[
\rho_{\mathcal{E}_{A^{\prime}\rightarrow B_{1}^{\prime}}}^{\mu}\otimes
\bigotimes\nolimits_{i=1}^{M}\rho_{E_{i}}(\bar{n}_{i})\right]  , \label{ff1}%
\end{equation}
where $\rho_{\mathcal{E}_{A^{\prime}\rightarrow B_{1}^{\prime}}}^{\mu
}:=\mathcal{I}_{A}\otimes\mathcal{E}_{A^{\prime}\rightarrow B_{1}^{\prime}%
}(\Phi_{AA^{\prime}}^{\mu})$ is associated with the first beamsplitter, and
$\mathbf{L}_{A|E_{1}\cdots E_{M}}$ is a trace-preserving LOCC, local with
respect to the cut $A|B_{1}^{\prime}E_{1}\cdots E_{M}$. Also note that, for
any separable state $\sigma_{s}^{\mu}(A|B_{1}^{\prime})$ we have that the
output state%
\begin{equation}
\tilde{\sigma}_{s}^{\mu}=\mathbf{L}_{A|B_{1}^{\prime}E_{1}\cdots E_{M}}\left[
\sigma_{s}^{\mu}(A|B_{1}^{\prime})\otimes\bigotimes\nolimits_{i=1}^{M}%
\rho_{E_{i}}(\bar{n}_{i})\right]  \label{ff2}%
\end{equation}
is separable with respect to the cut $A|B_{1}\cdots B_{M}$. As a result we
have that%
\begin{gather}
E_{R(A|B_{1}\cdots B_{M})}(\rho_{\mathcal{E}})\overset{(1)}{\leq}\inf
_{\tilde{\sigma}_{s}^{\mu}(A|B_{1}\cdots B_{M})}\underset{\mu\rightarrow
+\infty}{\lim\inf}S(\rho_{\mathcal{E}}^{\mu}||\tilde{\sigma}_{s}^{\mu
})\nonumber\\
\overset{(2)}{\leq}\inf_{\sigma_{s}^{\mu}(A|B_{1}^{\prime})}\underset
{\mu\rightarrow+\infty}{\lim\inf}S(\rho_{\mathcal{E}_{A^{\prime}\rightarrow
B_{1}^{\prime}}}^{\mu}||\sigma_{s}^{\mu}):=\Phi(\mathcal{E}_{A^{\prime
}\rightarrow B_{1}^{\prime}}),
\end{gather}
where we use: (1) the fact that $\tilde{\sigma}_{s}^{\mu}(A|B_{1}\cdots
B_{M})$ are specific types of $\sigma_{s}^{\mu}(A|B_{1}\cdots B_{M})$; and (2)
monotonicity and additivity of the relative entropy with respect to the
decompositions in Eqs.~(\ref{ff1}) and~(\ref{ff2}).

Because $\mathcal{E}_{A^{\prime}\rightarrow B_{1}^{\prime}}$ is a thermal-loss
channel with transmissivity $\eta_{0}$ and mean photon number $\bar{n}_{0}$,
its entanglement flux is bounded by~\cite{PLOB}%
\begin{equation}
\Phi(\mathcal{E}_{A^{\prime}\rightarrow B_{1}^{\prime}})\leq-\log_{2}\left[
(1-\eta_{0})\eta_{0}^{\bar{n}_{0}}\right]  -h(\bar{n}_{0}), \label{LossUB}%
\end{equation}
for $\bar{n}_{0}<\eta_{0}/(1-\eta_{0})$, while zero otherwise. Here we set
\begin{equation}
h(x):=(x+1)\log_{2}(x+1)-x\log_{2}x. \label{hEntropyMAIN}%
\end{equation}
Thus, we find that the capacity between Alice and the $i$th Bob must satisfy%
\begin{equation}
\mathcal{C}^{i}\leq\left\{
\begin{array}
[c]{c}%
-\log_{2}\left[  (1-\eta_{0})\eta_{0}^{\bar{n}_{0}}\right]  -h(\bar{n}%
_{0})~~~~\text{for~~}\bar{n}_{0}<\frac{\eta_{0}}{1-\eta_{0}},\\
\\
0~~~~\text{for~~}\bar{n}_{0}\geq\frac{\eta_{0}}{1-\eta_{0}}%
.~~~~~~~~~~~~~~~~~~~~~~~~~~~~~~~~~
\end{array}
\right.  \label{Cithermal}%
\end{equation}
As expected, the first beamsplitter is a universal bottleneck which restricts
the capacities between Alice and any of the receiving Bobs.

In the specific case of a lossy broadcast channel with no thermal noise
($n_{i}=0$ for any $i$), we may specify Eq.~(\ref{Cithermal}) into the
following simple bound%
\begin{equation}
\mathcal{C}^{i}\leq-\log_{2}(1-\eta_{0})~.
\end{equation}
Let us note that, contrary to another work~\cite{Wilde} also inspired by
Ref.~\cite{PLOB}, our analysis of the lossy broadcast channel builds upon a
rigorous extension of channel simulation and teleportation stretching to CV
systems, which includes a suitable generalization of the REE to asymptotic
states. Since our results represent a\ rigorous extension of Ref.~\cite{PLOB},
they may also be used to solidify the claims presented in Ref.~\cite{Wilde} on
the capacity region of the lossy broadcast channel. For further details see
Ref.~\cite{TQCreview}.

Most importantly, notice that our derivation can be generalized to other
bosonic broadcast channels, where the $M+1$ beamsplitters are replaced by
arbitrary Gaussian unitaries $U_{A^{\prime}E_{0}},U_{B_{1}^{\prime}E_{1}%
},\ldots,U_{B_{M}^{\prime}E_{M}}$. In this general case, we repeat the
previous reasonings to find that the capacities must satisfy\ the bottleneck
relation
\begin{equation}
\mathcal{C}^{i}\leq\Phi(\mathcal{E}_{A^{\prime}\rightarrow B_{1}^{\prime}}),
\end{equation}
where the latter is the entanglement flux of the first Gaussian channel
$\mathcal{E}_{A^{\prime}\rightarrow B_{1}^{\prime}}$, determined by the action
of the Gaussian unitary $U_{A^{\prime}E_{0}}$ on the input mode $A^{\prime}$
and the thermal mode $E_{0}$.

\section{Quantum multiple-access channel\label{SECmulti}}

Let us now study multipoint-to-point quantum communication, i.e., a quantum
multiple-access channel from $M$ senders (Alices) to a single receiver (Bob).
This channel is a CPTP map from Alices' input space $\mathcal{D}(\otimes
_{i}\mathcal{H}_{a^{i}})$ to Bob's output space $\mathcal{D}(\mathcal{H}_{b}%
)$. The most general adaptive protocol over this channel goes as follows. All
the parties prepare their initial systems by means of a LOCC $\Lambda_{0}$.
Then, the $i$th Alice picks the first system from her local ensemble, i.e.,
$a_{1}^{i}\in\mathbf{a}^{i}$. All Alice's input systems are sent through the
quantum multiple-access channel $\mathcal{E}$ with output $b_{1}$ for Bob,
i.e.,
\begin{equation}
a_{1}^{1},\ldots,a_{1}^{i},\ldots,a_{1}^{M}\overset{\mathcal{E}}{\rightarrow
}b_{1}~.
\end{equation}
This is followed by another LOCC $\Lambda_{1}$ involving all parties. Bob
ensemble is updated as $b_{1}\mathbf{b}\rightarrow\mathbf{b}$. Then, there is
the second transmission $\{\mathbf{a}^{i}\}\ni\{a_{2}^{i}\}\rightarrow b_{2}$
through $\mathcal{E}$, followed by another LOCC $\Lambda_{2}$ and so on. After
$n$ uses, the $i$th Alice and Bob share an output state $\rho_{\mathbf{a}%
^{i}\mathbf{b}}^{n}$ which is epsilon-close to a target state with $nR_{i}%
^{n}$ bits.

The generic multiple-access capacity for the $i$th Alice is defined by
maximizing the asymptotic rate over all the adaptive LOCCs $\mathcal{L}%
=\{\Lambda_{0},\Lambda_{1},\ldots\}$, i.e., we have $\mathcal{C}^{i}%
:=\sup_{\mathcal{L}}\lim_{n}R_{i}^{n}$. As before, by specifying the adaptive
protocol to a particular task, one derives the entanglement distribution
multiple-access capacity ($D_{2}^{i}$), the quantum multiple-access capacity
($Q_{2}^{i}$), the secret-key multiple-access capacity ($K^{i}$) and the
private multiple-access capacity ($P_{2}^{i}$). These are all assisted by
unlimited two-way CCs between the parties and satisfy $D_{2}^{i}=Q_{2}^{i}\leq
K^{i}=P_{2}^{i}$.

\begin{figure}[ptbh]
\vspace{-1.7cm}
\par
\begin{center}
\includegraphics[width=0.48\textwidth]{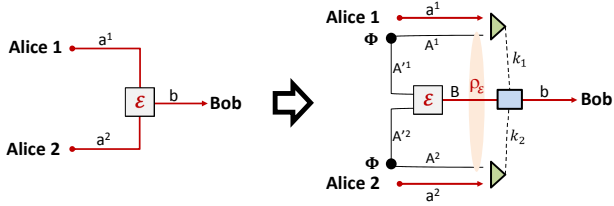} \vspace{0.3cm} \vspace
{-2.7cm}
\end{center}
\caption{Simulation of a teleportation-covariant quantum multiple-access
channel. We can replace the multiple-access channel $\mathcal{E}:a^{1}%
a^{2}\rightarrow b$ (left)\ by double teleportation over its tripartite Choi
matrix $\rho_{\mathcal{E}}$ (right). This Choi matrix is obtained by sending
halves ($A^{\prime1}$ and $A^{\prime2}$) of two EPR\ states $\Phi$\ through
$\mathcal{E}$, with output $B$. Then, systems $a^{1}$ and $A^{1}$ are subject
to a Bell detection with outcome $k_{1}$. Similarly, systems $a^{2}$ and
$A^{2}$ are subject to a Bell detection with outcome $k_{2}$. The outcomes are
CCed to Bob who applies a correction unitary on system $B$. Since the channel
is teleportation-covariant, i.e., it commutes with the teleportation unitaries
according to Eq.~(\ref{stretchMULTI}), Bob's correction unitary $V_{k}^{-1}$
on $B$ re-generates the original channel $\mathcal{E}:a^{1}a^{2}\rightarrow
b$.}%
\label{multiple}%
\end{figure}

Let us introduce the notion of teleportation-covariant multiple-access
channel. For the sake of simplicity, this is explained for the case of two
senders, with the extension to arbitrary $M$ senders being just a matter of
technicalities. We also consider the case of DV\ channels, with the extension
to CV\ channels left implicit and following the basic methodology of
Sec.~\ref{review}. A quantum multiple-access channel is
teleportation-covariant if, for any teleportation unitaries, $U_{k_{1}}^{1}$
and $U_{k_{2}}^{2}$, we may write
\begin{equation}
\mathcal{E}\left[  (U_{k_{1}}^{1}\otimes U_{k_{2}}^{2})\rho(U_{k_{1}}%
^{1}\otimes U_{k_{2}}^{2})^{\dagger}\right]  =V_{k}\mathcal{E}(\rho
)V_{k}^{\dagger}, \label{stretchMULTI}%
\end{equation}
for some unitary $V_{k}$, with $k$ depending on both $k_{1}$ and $k_{2}$. If
this is the case, then we can replace $\mathcal{E}$ with teleportation over
its Choi matrix, which is defined by sending halves of two EPR states through
the channel, i.e.,%
\begin{equation}
\rho_{\mathcal{E}}=\mathcal{I}_{A^{1}A^{2}}\otimes\mathcal{E}_{A^{\prime
1}A^{\prime2}}(\Phi_{A^{1}A^{\prime1}}\otimes\Phi_{A^{2}A^{\prime2}}).
\label{choi}%
\end{equation}
See also Fig.~\ref{multiple} for further explanations.

\begin{figure}[ptbh]
\vspace{-2.3cm}
\par
\begin{center}
\vspace{-0.3cm} \vspace{0.8cm}
\includegraphics[width=0.55\textwidth]{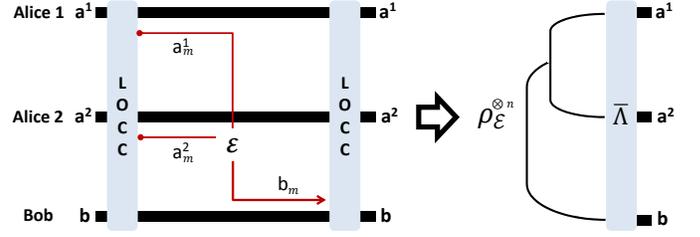} \vspace{1.1cm}
\vspace{-3.5cm}
\end{center}
\caption{Teleportation stretching of an adaptive protocol implemented over a
teleportation-covariant multiple-access channel (generic $m$th transmission
shown on the left). After $n$ uses, we can express the output in terms of $n$
copies of the Choi matrix $\rho_{\mathcal{E}}$ of the quantum multiple-access
channel, subject to a trace-preserving LOCC $\bar{\Lambda}$.}%
\label{mutlpleST}%
\end{figure}

By using the channel simulation, we may fully simplify any adaptive protocol
performed over a teleportation-covariant multiple-access channel $\mathcal{E}%
$. In fact, each transmission through $\mathcal{E}$ can be replaced by double
teleportation on its Choi matrix $\rho_{\mathcal{E}}$, with the Bell
detections and Bob's correction unitary being included in the LOCCs of the
protocol. By stretching $n$ uses of the adaptive protocol (see
Fig.~\ref{mutlpleST}), we find that the total output state of Alice 1, Alice 2
and Bob reads%
\begin{equation}
\rho_{\mathbf{a}^{1}\mathbf{a}^{2}\mathbf{b}}^{n}=\bar{\Lambda}\left(
\rho_{\mathcal{E}}^{\otimes n}\right)  .
\end{equation}
If we now trace one of the two senders, e.g., Alice 2, we still have an LOCC
between Alice 1 and Bob. In other words, we may write the following
\begin{equation}
\rho_{\mathbf{a}^{1}\mathbf{b}}^{n}=\bar{\Lambda}_{\mathbf{a}^{1}%
\mathbf{a}^{2}|\mathbf{b}}\left(  \rho_{\mathcal{E}}^{\otimes n}\right)  ,
\end{equation}
where $\bar{\Lambda}_{\mathbf{a}^{1}\mathbf{a}^{2}|\mathbf{b}}$ is local with
respect to the cut $\mathbf{a}^{1}\mathbf{a}^{2}|\mathbf{b}$%
.

For Alice 1 and Bob ($i=1$), we can now write%
\begin{align}
E_{R}(\rho_{\mathbf{a}^{1}\mathbf{b}}^{n})  &  :=\inf_{\sigma_{s}%
(\mathbf{a}^{1}|\mathbf{b})}S\left(  \rho_{\mathbf{a}^{1}\mathbf{b}}%
^{n}||\sigma_{s}\right) \nonumber\\
&  \leq\inf_{\sigma_{s}(\mathbf{a}^{1}\mathbf{a}^{2}|\mathbf{b})}S\left(
\rho_{\mathcal{E}}^{\otimes n}||\sigma_{s}\right)  :=E_{R(\mathbf{a}%
^{1}\mathbf{a}^{2}|\mathbf{b})}(\rho_{\mathcal{E}}^{\otimes n})\nonumber\\
&  \leq\inf_{\sigma_{s}(\mathbf{a}^{1}|\mathbf{a}^{2}|\mathbf{b})}S\left(
\rho_{\mathcal{E}}^{\otimes n}||\sigma_{s}\right)  :=E_{R}(\rho_{\mathcal{E}%
}^{\otimes n}).
\end{align}
By applying the weak converse bound, we then derive%
\begin{equation}
\mathcal{C}^{1}\leq\sup_{\mathcal{L}}\underset{n}{\lim}\frac{E_{R}%
(\rho_{\mathbf{a}^{1}\mathbf{b}}^{n})}{n}\leq E_{R(\mathbf{a}^{1}%
\mathbf{a}^{2}|\mathbf{b})}^{\infty}(\rho_{\mathcal{E}})\leq E_{R}^{\infty
}(\rho_{\mathcal{E}}),
\end{equation}
and using the subadditivity of the REE over tensor products, it is easy to
show the single-letter version
\begin{equation}
\mathcal{C}^{1}\leq E_{R(\mathbf{a}^{1}\mathbf{a}^{2}|\mathbf{b})}%
(\rho_{\mathcal{E}})\leq E_{R}(\rho_{\mathcal{E}}).
\end{equation}

Note that we find the same bound for the other capacity for Alice 2 and Bob
($i=2$). The reasoning can be readily extended to arbitrary $M$ senders, so
that the capacity between the $i$th Alice and Bob reads%
\begin{equation}
\mathcal{C}^{i}\leq E_{R(\mathbf{a}^{1}\cdots\mathbf{a}^{M}|\mathbf{b})}%
(\rho_{\mathcal{E}})\leq E_{R}(\rho_{\mathcal{E}}):=\Phi(\mathcal{E}),
\end{equation}
where $\Phi(\mathcal{E})$ is the entanglement flux of the quantum
multiple-access channel. As previously mentioned, the result can be extended
to CV systems by employing asymptotic simulations and extending the notions.



\section{All-in-all quantum communication\label{SECall}}

In this section we extend our technique to a single-hop quantum network
involving multiple ($M_{A}$) senders and multiple ($M_{B}$)\ receivers, which
is also known as quantum interference channel. This is a CPTP map from Alices'
input space $\mathcal{D}(\otimes_{i=1}^{M_{A}}\mathcal{H}_{a^{i}})$ to Bobs'
output space $\mathcal{D}(\otimes_{j=1}^{M_{B}}\mathcal{H}_{b^{j}})$. As a
straightforward generalization of the previous cases, the most general
adaptive protocol over this channel can be described as follows. At the
initial stage the parties exploit a LOCC $\Lambda_{0}$ for their systems'
preparation. Then, each Alice picks the first system from her local ensemble
$a_{1}^{i}\in\mathbf{a}^{i}$. The inputs of all Alices are sent to all Bobs
through channel $\mathcal{E}$ resulting into the outputs $\{b_{1}^{i}\}$,
i.e.,%
\begin{equation}
a_{1}^{1},\ldots,a_{1}^{i},\ldots,a_{1}^{M_{A}}\overset{\mathcal{E}%
}{\rightarrow}b_{1}^{1},\ldots,b_{1}^{j},\ldots,b_{1}^{M_{B}}~.
\end{equation}
After this first transmission, there is another LOCC $\Lambda_{1}$, after
which all Bobs' ensembles are updated $b_{1}^{j}\mathbf{b}^{j}\rightarrow
\mathbf{b}^{j}$. Next, there is the second transmission $\mathbf{a}^{i}%
\ni\{a_{2}^{i}\}\rightarrow\{b_{2}^{j}\}$ through $\mathcal{E}$, followed by
another LOCC $\Lambda_{2}$ and so on.

Thus, after $n$ uses of the channel, the $i$th Alice and the $j$th Bob share
an output state $\rho_{\mathbf{a}^{i}\mathbf{b}^{j}}^{n}$, which is $\epsilon
$-close to a target state of $nR_{ij}^{n}$ bits. By maximizing the asymptotic
rate over all the adaptive LOCCs $\mathcal{L}=\{\Lambda_{0},\Lambda_{1}%
,\ldots\}$ we can define the generic interference capacity for the $i$th Alice
and the $j$th Bob as
\begin{equation}
\mathcal{C}^{ij}:=\sup_{\mathcal{L}}\lim_{n}R_{ij}^{n}~.
\end{equation}
As usual, depending on the task, one specifies three different capacities
assisted by unlimited two-way CCs: The entanglement distribution capacity
($D_{2}^{ij}$), the quantum capacity ($Q_{2}^{ij}$), the secret-key capacity
($K^{ij}$) and the private capacity ($P_{2}^{ij}$)\ of the quantum
interference channel (with $D_{2}^{ij}=Q_{2}^{ij}\leq K^{ij}=P_{2}^{ij}$).

\begin{figure}[ptbh]
\vspace{-1.5cm}
\par
\begin{center}
\vspace{-0.5cm} \vspace{0.5cm} \vspace{-1.2cm}
\includegraphics[width=0.65\textwidth]{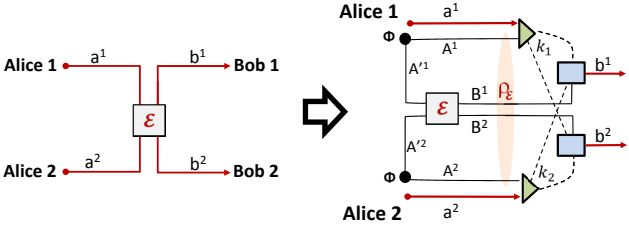} \vspace{-1.0cm}
\vspace{-2.3cm}
\end{center}
\caption{Simulation of a teleportation-covariant quantum interference channel.
The channel $\mathcal{E}:a^{1}a^{2}\rightarrow b^{1}b^{2}$ (left) can be
simulated by its Choi matrix $\rho_{\mathcal{E}}$ (right). Systems $a^{1}$ and
$A^{1}$ are subject to a Bell detection with outcome $k_{1}$. Similarly,
systems $a^{2}$ and $A^{2}$ are subject to a Bell detection with outcome
$k_{2}$. Both outcomes $k_{1}$ and $k_{2}$ are then classically communicated
to Bob 1 and Bob 2 who apply two correction unitaries on $B^{1}$ and $B^{2}$.
Since the channel is teleportation-covariant, i.e., it commutes with the
teleportation unitaries according to Eq.~(\ref{allSTRE}), the two Bobs are
able to recover the original channel $\mathcal{E}:a^{1}a^{2}\rightarrow
b^{1}b^{2}$ by applying correction unitaries $(V_{l_{1}}^{1})^{-1}$ and
$(V_{l_{2}}^{2})^{-1}$.}%
\label{bus1}%
\end{figure}

As for the case of the broadcast and the multiple-access channels we bound
these capacities by using REE+teleportation stretching. We proceed by
considering two senders and two receivers being the extension to arbitrary
$M_{A}$ and $M_{B}$ just a matter of technicalities. The definition of a
teleportation-covariance quantum interference channel relies once again on the
commutation with teleportation, i.e., for any teleportation unitaries
$U_{k_{1}}^{1}$ and $U_{k_{2}}^{2}$ we must have
\begin{equation}
\mathcal{E}\left[  (U_{k_{1}}^{1}\otimes U_{k_{2}}^{2})\rho(U_{k_{1}}%
^{1}\otimes U_{k_{2}}^{2})^{\dagger}\right]  =\mathcal{V}\mathcal{E}%
(\rho)\mathcal{V}^{\dagger}, \label{allSTRE}%
\end{equation}
where $\mathcal{V}=V_{l_{1}}^{1}\otimes V_{l_{2}}^{2}$ for unitaries
$V_{l_{1}}^{1}$ and $V_{l_{2}}^{2}$, with both $l_{1}$ and $l_{2}$ depending
on $k_{1}$ and $k_{2}$. If this condition holds then the channel can be
simulated by teleportation over its Choi matrix, which is formally defined as
in Eq.~(\ref{choi}). See Fig.~\ref{bus1}\ for this
simulation.

\begin{figure}[ptbh]
\vspace{-2.1cm}
\par
\begin{center}
\includegraphics[width=0.49\textwidth]{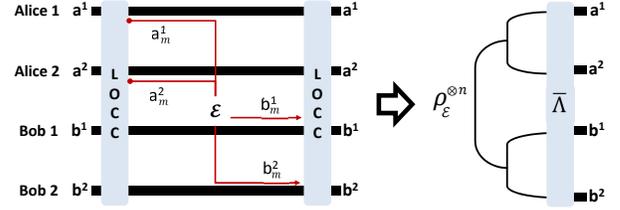} \vspace{-2.3cm}
\end{center}
\caption{Teleportation stretching of an adaptive protocol over a quantum
interference channel (generic $m$th transmission shown on the left). After $n$
uses, we can express the output in terms of $n$ copies of the Choi matrix
$\rho_{\mathcal{E}}$ of the quantum interference channel, subject to a
trace-preserving LOCC $\bar{\Lambda}$.}%
\label{bus2}%
\end{figure}

Thus, an adaptive protocol can be simplified since each use of channel
$\mathcal{E}$ can be replaced by teleportation and both the Bell detections
and Bobs' correction unitaries become part of the LOCCs. By stretching $n$
uses of the channel (see Fig.~\ref{bus2}), we have the following output state
shared between Alice~1, Alice~2, Bob~1 and Bob~2
\begin{equation}
\rho_{\mathbf{a}^{1}\mathbf{a}^{2}\mathbf{b}^{1}\mathbf{b}^{2}}=\bar{\Lambda
}(\rho_{\mathcal{E}}^{\otimes n}).
\end{equation}
By tracing over one sender and one receiver, say Alice~2 and Bob~2, we then
derive
\begin{equation}
\rho_{\mathbf{a}^{1}\mathbf{b}^{1}}^{n}=\bar{\Lambda}_{\mathbf{a}%
^{1}\mathbf{a}^{2}|\mathbf{b}^{1}\mathbf{b}^{2}}(\rho_{\mathcal{E}}^{\otimes
n})~,
\end{equation}
where $\bar{\Lambda}_{\mathbf{a}^{1}\mathbf{a}^{2}|\mathbf{b}^{1}%
\mathbf{b}^{2}}$ is a trace-preserving LOCC between Alice~1 and Bob~1, local
with respect to the cut $\mathbf{a}^{1}\mathbf{a}^{2}|\mathbf{b}^{1}%
\mathbf{b}^{2}$. 

It follows that the capacity for Alice 1 and Bob 1 ($i=j=1$) is upper bounded by%

\begin{equation}
\mathcal{C}^{11}\leq\sup_{\mathcal{L}}\underset{n}{\lim}\frac{E_{R}%
(\rho_{\mathbf{a}^{1}\mathbf{b}^{1}}^{n})}{n}\leq E_{R(\mathbf{a}%
^{1}\mathbf{a}^{2}|\mathbf{b}^{1}\mathbf{b}^{2})}^{\infty}(\rho_{\mathcal{E}%
})\leq E_{R}^{\infty}(\rho_{\mathcal{E}}).
\end{equation}
In terms of single-letter bounds we find%
\begin{equation}
\mathcal{C}^{11}\leq E_{R(\mathbf{a}^{1}\mathbf{a}^{2}|\mathbf{b}%
^{1}\mathbf{b}^{2})}(\rho_{\mathcal{E}})\leq E_{R}(\rho_{\mathcal{E}}).
\end{equation}
Clearly, we find the same result in all other cases, i.e., for any
sender-receiver pair $(i,j)$. In general, for arbitrary $M_{A}$ senders and
$M_{B}$\ receivers, we may write%
\begin{equation}
\mathcal{C}^{ij}\leq E_{R(\mathbf{a}^{1}\cdots\mathbf{a}^{M_{A}}%
|\mathbf{b}^{1}\cdots\mathbf{b}^{M_{B}})}(\rho_{\mathcal{E}})\leq E_{R}%
(\rho_{\mathcal{E}}):=\Phi(\mathcal{E}),
\end{equation}
where $\Phi(\mathcal{E})$\ is the entanglement flux of the quantum
interference channel. The extension to CV systems exploits asymptotic
simulations along the lines of Sec.~\ref{review}.

\section{Conclusions\label{SECconclu}}

In this work we have studied the capacities for quantum communication,
entanglement distribution and secret key generation in a single-hop quantum
network, involving a direct channel between multiple senders and/or multiple
receivers. More precisely, we have considered the quantum broadcast channel
(point-to-multipoint), the multiple-access channel (multipoint-to-point), and
the quantum interference channel (multipoint-to-multipoint), assuming that all
the parties may apply the most general local operations assisted by unlimited
two-way CCs (adaptive protocols).

By suitably extending the\ methodology of Ref.~\cite{PLOB}, which suitably
combines the relative entropy of entanglement (REE) with teleportation
stretching, we have reduced the most general adaptive protocols implemented on
these multipoint channels to the computation of a one-shot quantity and, in
particular, their entanglement flux (i.e., the REE of their Choi matrix). This
is achieved at any dimension, i.e., finite or infinite (CV channels).

Further research should be directed to show how a rigorous application of
our\ reduction method can be used to upperbound the entire capacity regions of
multipoint channels, defined as the convex closure of the set of all the rates
which are achievable by the parties assisted by unlimited two-way CCs. Other
important directions are related with the study of multipoint quantum
communication within a multi-hop quantum network, following the general
methods and results of Ref.~\cite{networkPIRS}.

\bigskip

\textbf{Acknowledgments}. This work has been supported by the EPSRC via the
`UK Quantum Communications Hub' (EP/M013472/1). The authors would like to
thank S. L. Braunstein, S. Lloyd, G. Spedalieri, and C. Ottaviani for valuable
feedback and comments.


\begin{thebibliography}{99}
\bibitem {Eps}More precisely, this defines an $(n,R^{n},\varepsilon
)$-protocol, but we omit this\ technical notation here in order to simplify
the physical discussion.
\bibitem {monotonocity}Clearly, this step can be broken in two parts. First,
the monotonicity of the REE under partial trace and then under the
trace-preserving LOCC $\bar{\Lambda}$, so that $E_{R}(\rho_{\mathbf{ab}}%
^{n})\leq E_{R(\mathbf{a}|\mathbf{bc})}(\rho_{\mathbf{abc}}^{n})\leq
E_{R(\mathbf{a}|\mathbf{bc})}(\rho_{\mathcal{E}}^{\otimes n})$.
\end{thebibliography}
\end{document}